\documentclass[12pt,a4paper]{article}

\usepackage[a4paper,left=2.1cm,right=2.1cm,bottom=2.2cm,top=2.5cm,footskip=32pt]{geometry}
\usepackage{siunitx}
\usepackage{pifont}
\newcommand{\xmark}{\ding{55}}
\usepackage{setspace}
\usepackage{time}
\usepackage[sort]{natbib}
\usepackage{graphicx}
\usepackage{amsmath,amssymb,bm}
\usepackage[table,xcdraw,dvipsnames]{xcolor}
\usepackage{hyperref}
\usepackage{multirow}
\usepackage{cleveref}
\usepackage{rotating}
\usepackage{authblk}
\usepackage{longtable}
\usepackage{booktabs}
\usepackage{tikz}
\usetikzlibrary{arrows.meta, positioning}
\usetikzlibrary{shapes.arrows}

\usepackage{listings}

\definecolor{codegreen}{rgb}{0,0.6,0}
\definecolor{codegray}{rgb}{0.5,0.5,0.5}
\definecolor{codepurple}{rgb}{0.58,0,0.82}
\definecolor{backcolour}{rgb}{0.95,0.95,0.92}

\lstdefinestyle{mystyle}{
    backgroundcolor=\color{backcolour},  
    commentstyle=\color{codegreen},
    keywordstyle=\color{magenta},
    numberstyle=\color{white},
    stringstyle=\color{codepurple},
    basicstyle=\ttfamily\small,
    breakatwhitespace=false,         
    breaklines=true,    
    captionpos=b,                    
    keepspaces=true, 
    numbers=left,                    
    numbersep=5pt, 
    showspaces=false,                
    showstringspaces=false, 
    showtabs=false, 
    tabsize=2}
\newsavebox\CBox

\newcommand\myshade{85}
\colorlet{mylinkcolor}{violet}
\colorlet{mycitecolor}{YellowOrange}
\colorlet{myurlcolor}{Aquamarine}

\hypersetup{
    linkcolor  = mylinkcolor!\myshade!black,
    citecolor  = mycitecolor!\myshade!black,
    urlcolor   = myurlcolor!\myshade!black,
    colorlinks = true,
}
\usepackage{subfig}

\begin{document}

\title{\Huge{Fast Estimation of the Composite Link Model for Multidimensional Grouped Counts}
}

\author[1]{Carlo G.~Camarda\footnote{Corresponding author: \url{carlo-giovanni.camarda@ined.fr}\\	\hspace*{1.8em}Address: 9 cours des Humanités, 93322 Aubervilliers - France}}
\author[2]{María Durbán}
\affil[1]{\small Institut National d'Études Démographiques, Aubervillers, France}	
\affil[2]{\small Department of Statistics, Universidad Carlos III de Madrid, Spain}

\date{}
\maketitle

\begin{abstract}
This paper presents a significant advancement in the estimation of the Composite Link Model within a penalized likelihood framework, specifically designed to address indirect observations of grouped count data. While the model is effective in these contexts, its application becomes computationally challenging in large, high-dimensional settings. To overcome this, we propose a reformulated iterative estimation procedure that leverages Generalized Linear Array Models, enabling the disaggregation and smooth estimation of latent distributions in multidimensional data. Through simulation studies and applications to high-dimensional mortality datasets, we demonstrate the model’s capability to capture fine-grained patterns while comparing its computational performance to the conventional algorithm. The proposed methodology offers notable improvements in computational speed, storage efficiency, and practical applicability, making it suitable for a wide range of fields in which high-dimensional data are provided in grouped formats.
\\$\,$
\end{abstract}	

\onehalfspacing

\section{Introduction}\label{sec:Intro}

The Composite Link Model (CLM) is an advanced framework designed in situations where it is necessary to link each observation with a linear function of more than one predicted value. This is the case when the aim is to  model indirect observations of counts, and this particular setting will be the focus of this paper. Initially proposed by \citet{ThompsonBakerCLM1981}, the CLM extends the capabilities of the Generalized Linear Model \citep[GLM,][]{McCullaghNelderGLM, NelderWedderburnGLM1972}. The CLM enables the incorporation of complex relationships and dependencies in count data, making it particularly useful in scenarios where direct observations are not feasible. This extension maintains the flexibility and interpretability of the GLM while addressing the unique challenges posed by indirect count data. 

A significant application of this methodology, and one that has led to the reformulation of the CLM proposed here, is in estimating the underlying or latent process behind observed grouped data. These grouped data can be considered indirect observations of the latent process, highlighting the CLM's ability to uncover the deeper structures within aggregated or incomplete datasets. 

In areas such as demography or epidemiology, this situation often arises when working with mortality tables, where death counts are aggregated over ages and/or years. Using this context as an illustration, the aim of this paper is to estimate the underlying latent mortality patterns across both ages and years, and potentially months and weeks. In the absence of prior knowledge, it is natural to exploit the inherent order in the data and impose smoothness to obtain unique estimates. This can be achieved by incorporating a roughness penalty into the associated likelihood function as described by \citet{Eilers2007}. Hence, throughout this paper, we refer to it as the Penalized Composite Link Model (PCLM). The resulting scoring algorithm is a modified version of the classic iterative reweighted least-squares (IRLS) for GLMs, incorporating an additional penalty and a revised design matrix to account for the grouping structure.  

While the PCLM has been successfully applied in similar contexts \citep{RizziEtAlPCLM1D2015, RizziEtAlPCLM2D2018}, including within Bayesian estimation frameworks \citep{LambertEilersBayesGrouped2009, LambertBayesBivariateCSDA2011}, the problem size in two- or higher-dimensional settings often becomes quite large, rendering the computational cost of maximizing the penalized likelihood function infeasible. Recently, deep learning approaches have been proposed for ungrouping data across multiple populations, a setting that can be viewed as moderately high-dimensional \citep{NigriEtAlDisaggregating2024}. These methods, however, are tailored to mortality data and lie outside the PCLM framework: they do not model the likelihood of the observed aggregated counts and therefore do not allow for likelihood-based inference. In contrast, this paper focuses on the PCLM itself and proposes an alternative estimation procedure that enables its efficient estimation in genuinely high-dimensional settings, based on a reformulation of IRLS that resembles the EM-algorithm \citep{DempsterEtAl_EM1977}.
We alternate between computation of the latent distribution of counts from the current parameter estimates and estimating the parameters from the actual and estimated observations. Although it is well established that the EM algorithm can suffer from slow convergence \citep{NgEtAl_EMchapter2012}, the computational and storage advantages of our approach stem from expressing the whole algorithm as a Generalized Linear Array Model \citep[GLAM,][]{CurrieDurbanEilersGLAM2006}. This formulation avoids the need to explicitly construct large and memory-intensive model matrices and efficiently handles the large systems of equations that arise. As a result, our method achieves substantial gains in both memory usage and computational speed. Furthermore, unlike the EM algorithm, which assumes a Poisson distribution for latent variables rather than for the observed aggregated counts, our approach preserves the original probabilistic assumptions of the model. It does not alter the underlying data-generating process, but instead constitutes a re-engineering of the estimation procedure to accommodate high-dimensional settings efficiently.

This methodology could be extremely useful in common situations where the grouping structure in one dimension is constant across the other dimension (this is a common case in mortality tables where we observed aggregated death counts in coarse age groups and the aggregation structure is similar for all years and other dimensions).  Although exact uniform grouping may not always be observed in real-world datasets, the assumption of consistent aggregation is often approximately satisfied in many practical contexts, particularly in demography and spatio-temporal epidemiology. For instance, in spatio-temporal health surveillance, data are commonly collected at regular temporal intervals (e.g., weekly counts), even when spatial granularity is heterogeneous or sparse. In such cases, the temporal dimension typically imposes a structured and regular grouping pattern \citep{Leeetal2022}. Furthermore, aggregation structures in space-time settings often exhibit a regular and partially separable structure across dimensions. This characteristic enables them to be effectively represented using combinations of their marginal temporal and spatial components that are well aligned with the assumptions of our proposed modeling framework. As a result, the methodology retains broad applicability, particularly in settings where at least one dimension exhibits consistent and well-defined grouping patterns.

In summary, the main contributions of this work can be summarized in five points. First, a new computational formulation for Penalized Composite Link Models is introduced, which avoids the explicit construction of high-dimensional Kronecker-product design and composition matrices, a central feature of existing implementations. Second, this formulation enables the estimation of PCLMs in two or more dimensions at fine resolutions that are computationally infeasible or unattainable with currently available methods. Third, the original probabilistic assumptions of the Composite Link Model are fully preserved, avoiding the additional modeling assumptions and likelihood modifications inherent to EM-based approaches. Fourth, it allows full variance and uncertainty estimation in multidimensional settings, where existing algorithms frequently fail due to prohibitive memory requirements. Finally, the approach achieves substantial gains in computational speed and memory efficiency without compromising estimation accuracy.

The structure of this paper is as follows: following this Introduction, Section \ref{sec:CLM} provides an overview of the Penalized Composite Link Model, focusing on the new estimation methodology within the Generalized Linear Array Models framework and on methods for computing variance and standard errors. Section \ref{sec:Simulations} presents two simulation studies designed to evaluate the accuracy of the PCLM in recovering the latent distribution and to compare the computational performance of the proposed method against the original formulation. Section \ref{sec:Applications} illustrates the practical utility of the proposed approach through the analysis of two mortality datasets in two-dimensional and three-dimensional settings. Finally, Section \ref{sec:Conclusions} concludes by summarizing the main findings.

\section{The Penalized Composite Link Model over multidimensional arrays}\label{sec:CLM}

For clarity and simplicity, this paper primarily focuses on presenting the mathematical details for the two-dimensional case. However, the proposed methodology and framework can be easily extended to higher-dimensional settings, enabling the analysis of more complex data structures with minimal modifications, as briefly discussed in Section \ref{sec:GLAMandCLM}. Additionally, while the first example illustrates our approach using two-dimensional mortality data grouped by age and year, the second example addresses a large-scale regression problem in three dimensions, incorporating mortality data by age groups, years, and weeks.

To facilitate the discussion, we introduce the following key notation: $\bm{1}_{n}$ represents a column vector of ones of length $n$ while $\bm{1}_{m \times n}$ is a matrix of ones of size $m\times n$ and $\bm{I}_{n}$ denotes the identity matrix of size $n$. Additionally, we will frequently use the Kronecker product, denoted by $\otimes$ as well as both element-wise multiplication and element-wise division, denoted by $\odot$ and $\oslash$, respectively. Furthermore, let $\verb"vec"(\bm{M})$ represent the function that vectorizes matrix $\bm{M}$ by stacking its columns into a single vector. Lastly, $\verb"diag"(\bm{v})$ is the function that constructs a diagonal matrix with the elements of vector $\bm{v}$ positioned along its main diagonal.

Suppose that we observe an array of counts $\bm{Y} = (y_{i,j})$ (of size $n_1\times n_2$), where $n_1$ and $n_2$ correspond to the number of coarse-level observations along two covariates: $\bm{x}_{1} = (x_{11}, \ldots, x_{1n_1})$ and $\bm{x}_{2} = (x_{21}, \ldots, x_{2n_2})$ representing the grid over which the observations are aggregated. We assume that $Y_{i,j}$, the random variable corresponding to the observed counts $y_{i,j}$, follows the Poisson distribution with mean $\mu_{i,j}$, i.e.~ $Y_{i,j} \sim \mathcal{P}(\mu_{i,j})$. Let $\bm{y}=\verb"vec"(\bm{Y})$ denote the vectorized form of the observed counts (of length $n_1n_{2}$), and $\bm{\mu}$ represent the corresponding vector of expected values. In our setting, observed counts are the result of the contribution of several latent observations which we aim to estimate over a finer resolution of the covariates $\bm{x}_1$ and $\bm{x}_2$.  In other words, while we observe $\bm{y}$,  we aim to estimate the vector of latent observations $\bm{\gamma}=(\gamma_{11},\gamma_{12},\ldots, \gamma_{m_1m_2})$, where $m_1 > n_1$ and $m_2 > n_2$ denote the number of fine-scale levels along each covariate. The relation between the latent vector $\bm{\gamma}$ and $\bm{\mu}$ is specified by a known composition matrix $\bm{C}$  (of size $n_1n_2 \times m_1m_2$), which describes how the elements of the latent vector $\bm{\gamma}$ are combined to yield $\bm{\mu}$. The resulting Composite Link Model for Poisson aggregated counts can thus be written as follows:
\begin{equation} \label{eq:CLM}
   \bm{y} \sim \mathcal{P}(\bm{\mu}) \qquad \bm{\mu}=\bm{C}\bm{\gamma} \qquad \bm{\gamma}=\exp(\bm{\eta}),  
\end{equation}
where the structure of $\bm{C}$ will depend on the underlying process that generates the observed data, and $\bm{\eta}$ is the linear predictor in a Poisson framework.

For our illustrative example, we will utilize demographic data on mortality. Here, the counts represent observed deaths, with $\bm{x}_{1}$ corresponding to ages and $\bm{x}_{2}$ to calendar years. In many cases, especially when dealing with smaller geographic areas or specific causes of death, the available data are often aggregated over broader age groups or multiple years. Our objective is to estimate mortality at a more granular level, specifically for individual ages and single calendar years, to provide finer insights into mortality patterns. In general, it is common for the aggregation pattern in each dimension of the array to be consistent across all dimensions. For instance, if death counts are aggregated into 5-year age intervals, this grouping pattern is typically applied systematically across all (grouped) years of death. Furthermore, when analyzing death counts, it is essential to recognize that the expected values in the Poisson distribution are the product of the underlying force of mortality and the population at risk, commonly referred to as exposures: $\bm{y} \sim \mathcal{P}(\bm{e} \odot \bm{\mu})$. Exposures can be simply multiplied by the underlying latent force of mortality over a finer resolution since they are typically provided for single years of age and time, i.e.~$\bm{e}$ represents the vectorized form of the $m_{1} \times m_{2}$ matrix of observed exposures. These exposures serve as an offset in the Poisson regression framework, ensuring that mortality rates are accurately scaled relative to the population at risk. Moreover, exposures can be conveniently integrated into the algorithm at a later stage, preserving the overall structure of the computational framework.

The consistent aggregation structure results in a composition matrix of the form $\bm{C} = \bm{C}_2 \otimes \bm{C}_1$. Each marginal composition matrix $\bm{C}_d$, of dimension $n_d \times m_d$ for $d = 1, 2$, reflects the aggregation process in its respective dimension. This tensor product formulation captures the multidimensional nature of the aggregation, ensuring that the structural relationships are preserved across all dimensions. We also assume that each latent observation only contributes to one aggregated count, i.e.~the columns of the marginal composition matrices add up to one, e.g.~$\bm{1}_{m_{1}}' \bm{C}_1= \bm{1}'_{n_{1}}$; this is the case in the context of mortality tables and, as we will see in the next section, a key step in the formulation of the proposed approach as a Generalized Linear Array Model. 

Our objective is to describe the latent distribution 
$\bm{\gamma}$ by modeling the linear predictor $\bm{\eta}$. However, estimating a distribution with $m_1m_2$ values, which is significantly larger than the observed counts $n_1n_2$, requires imposing certain assumptions on the underlying latent structure. Given our limited understanding of this latent distribution, we propose that smoothness is a reasonable assumption for $\bm{\gamma}$, as suggested by \cite{Eilers2007}. A smooth curve inherently contains fewer details, which is appropriate given the scarcity of data. Unless there are sufficient data to support additional detailed features, smoothness allows us to extract useful information from data that might initially seem insufficient for providing any meaningful answers. 

Smoothing the latent distribution $\bm{\gamma}$ is equivalent to smoothing the linear predictor $\bm{\eta}$. Given the two-dimensional nature of the problem, we opted for a flexible, fully interactive bivariate smoothing function over $\bm{x}_1$ and $\bm{x}_2$:
\begin{equation}\label{eq:eta}
    \bm{\eta}=f(\bm{x}_1,\bm{x}_2)= \bm{B}\bm{\alpha}=(\bm{B}_2 \otimes \bm{B}_1)\bm{\alpha},
\end{equation}
where $\bm{B}_d$, $d=1,2$ are marginal $B$-spline bases over $\bm{x}_d$, of size $m_d\times c_d$, $d=1,2$. For a general reference on $B$-splines, see \cite{Boor2001}. \citet{CurrieDurbanEilersSmoothForecast2004} provides a clearer explanation of how to construct them in our specific context. The coefficients corresponding to each basis, denoted as $\bm{\alpha}$, must be estimated in order to derive the estimated latent distribution. 

The optimal level of smoothness can be achieved either by positioning $B$-splines strategically over the covariate space or by selecting an appropriate number of $B$-splines. Alternatively, following the approach introduced by \citet{EilersMarx1996} with $P$-splines, we decide to intentionally use a large number of equally-spaced $B$-splines to capture all relevant patterns in the data, then apply a penalty to remove unnecessary complexity, ensuring a parsimonious representation of the underlying structure. 
This translates to penalizing the coefficients $\bm{\alpha}$ in the resulting log-likelihood, a method widely employed for smoothing mortality over age and year \citep{CurrieDurbanEilersSmoothForecast2004, CamardaCPsplines2019} and when dealing with grouped death counts \citep{RizziEtAlPCLM2D2018}.

\subsection{Estimation procedure}\label{sec:estimation} 

The penalized log-likelihood for the Poisson PCLM in \eqref{eq:CLM}, where the linear predictor is defined as in \eqref{eq:eta}, was presented by \citet{Eilers2007} and holds for both one-dimensional and multidimensional cases. In both contexts, the core equation remains:
\begin{equation}\label{eq:loglik}
	\ell_P=\bm{y}' \ln\left[\bm{C} \exp(\bm{B}\bm{\alpha})\right]-\bm{1}_{n_1n_2}' \bm{C}  \exp(\bm{B}\bm{\alpha}) -\frac{1}{2}\bm{\alpha}' \bm{P} \bm{\alpha}\,. 
\end{equation}
The key distinctions in the multidimensional setting lie in three crucial components: the formulation of the $B$-spline basis matrix, the structure of the composition matrix $\bm{C}$, which we introduced in the previous section, and the penalty term $\bm{P}$. In particular, the penalty term $\bm{P}$, which smooths across both dimensions in the two-dimensional case, is expressed as:
\begin{equation}\label{eq:Pterm}
    \bm{P}=\lambda_1 (\bm{I}_{c_2}\otimes \bm{D}_1' \bm{D}_1)+\lambda_2(\bm{D}_2' \bm{D}_2 \otimes \bm{I}_{c_1})\, ,
\end{equation}
where $\lambda_1$ and $\lambda_2$ control the smoothness along each respective dimension, and $\bm{D}_d$ denotes the difference matrix for dimension  $d = 1, 2$ \citep{CurrieDurbanEilersSmoothForecast2004}. In the following, we use the second order differences. Different approaches can be employed to select the optimal combination of smoothing parameters $(\lambda_1, \lambda_2)$. However, in our context, the primary objective is to develop a more computationally efficient method for estimating a PCLM. As such, the choice of the specific selection criterion is secondary to our broader goals, and the robustness of our results will hold regardless of which criterion is ultimately chosen. We therefore subjectively select a specific combination of $(\lambda_1, \lambda_2)$ that demonstrates favorable performance in the illustrative applications discussed in Section \ref{sec:Applications}. Any further search for the optimal smoothing parameters would only scale the computational speed arithmetically, without affecting the relative computational efficiency between the conventional and proposed approaches.
 
As previously mentioned, we assume that the rows of $\bm{C}_d$, for $d = 1, 2$, do not overlap, ensuring that each latent observation contributes exclusively to a single grouped count. Consequently, the following identity holds:
\begin{equation}\label{eq:Crewritten}
\bm{1}_{n_1n_2}' \bm{C}=\bm{1}_{n_1n_2}' (\bm{C}_2 \otimes \bm{C}_1)= \verb"vec"(\bm{C}_2' \bm{1}_{n_2 \times n_1}\bm{C}_1)' = \verb"vec"(\bm{1}_{m_1 \times m_2})' =\bm{1}_{m_1m_2}',
\end{equation}
Then, \eqref{eq:loglik} can be rewritten as follows:
\begin{equation}\label{eq:loglik2}
	\ell_P=\bm{y}' \ln\left[\bm{C} \exp(\bm{B}\bm{\alpha})\right]-\bm{1}_{m_1m_2}'  \exp(\bm{B}\bm{\alpha}) -\frac{1}{2}\bm{\alpha}' \bm{P} \bm{\alpha},
\end{equation}
The derivative of \eqref{eq:loglik2} with respect to the vector $\bm{\alpha}$ is given by
\begin{equation}\label{eq:fd1}
	\frac{\partial \, \ell_P}{\partial\, \bm{\alpha}}= \bm{B}' \bm{\Gamma}\bm{C}' \bm{W}^{-1} \bm{y} - \bm{B}'  \exp(\bm{B}\bm{\alpha})- \bm{P} \bm{\alpha}\\
\end{equation}
where: $\bm{W}=\verb"diag"(\bm{\mu})$ and $\bm{\Gamma}=\verb"diag"(\bm{\gamma})$. In the original work by \citet{ThompsonBakerCLM1981}, and later extended by \cite{Eilers2007}, the solution to the system in equation \eqref{eq:fd1} was approached by defining a \emph{working matrix} $\breve{\bm{B}}=\bm{W}^{-1}  \bm{C} \bm{\Gamma} \bm{B}$ and solving the associated GLM using standard techniques. However, as the dimensionality of the data array increases, the explicit construction of matrix $\breve{\bm{B}}$ becomes computationally prohibitive, causing a significant escalation in memory and processing requirements.

To address this issue, we propose bypassing the direct computation of $\breve{\bm{B}}$ by introducing the concept of a \emph{working latent response} $\breve{\bm{y}}=\bm{\Gamma}\bm{C}' \bm{W}^{-1}\bm{y}$, which represents the redistribution of the observed counts at the desired resolution. In fact, this would correspond to the E-step in the EM algorithm 
 where, given current values of $\hat{\bm{\gamma}}$,  $\breve{y}_{k}=\frac{\hat \gamma_k}{\sum _l\hat \gamma_l}y_k$ would approximate the unobserved latent distribution. 
 
 As we will demonstrate in the following section, defining the working latent response enables the use of a Generalized Linear Array Model \citep[GLAM,][]{CurrieDurbanEilersGLAM2006}, which eliminates the need to compute $\breve{\bm{B}}$ or any Kronecker products involved in the estimation process. Consequently, our method is not a variant of the EM algorithm. Instead, it represents a computational strategy specifically designed to enable the estimation of a PCLM in a multidimensional framework.

Using the definition of the working latent response $\breve{\bm{y}}$, \eqref{eq:fd1} becomes:
\begin{equation}
\label{eq:fd11}
\frac{\partial\breve{\ell}_P}{\partial \bm{\alpha}} = \bm{B}' \breve{\bm{y}} -\bm{B}'  \exp(\bm{B}\bm{\alpha})-\bm{P} \bm{\alpha}\, .
\end{equation}

We now proceed with the standard Newton-Raphson approach to iteratively solve the system of equations
$$
\bm{0}=\frac{\partial \breve{\ell}_P}{\partial \bm{\alpha}} \approx \left. \frac{\partial \breve{\ell}_P}{\partial \bm{\alpha}}\right|_{\bm{\alpha} =\tilde{\bm{\alpha}} }
+ (\bm{\alpha}-\tilde{\bm{\alpha}}) 
\left. \frac{\partial^2 \breve{\ell}_P}{\partial \bm{\alpha}^2}\right|_{\bm{\alpha}=\tilde{\bm{\alpha}}}
$$
\begin{equation}\label{eq:NR}
\Rightarrow \quad \underbrace{\left. \frac{\partial^2 \breve{\ell}_P}{\partial \bm{\alpha}^2}\right|_{\bm{\alpha} = \tilde{\bm{\alpha}}} }_{\mbox{LHS}} \bm{\alpha} =
 \underbrace{\left. \frac{\partial^2 \breve{\ell}_P}{\partial \bm{\alpha}^2}\right|_{\bm{\alpha}=\tilde{\bm{\alpha}}} \tilde{\bm{\alpha}} - 
 \left. \frac{\partial \breve{\ell}_P}{\partial \bm{\alpha}}\right|_{\bm{\alpha}=\tilde{\bm{\alpha}}}}_{\mbox{RHS}}
\end{equation}
where LHS and RHS refer to the left- and right-hand sides of the iterative updates typical of the Newton-Raphson approach. 

To compute the LHS and RHS of equation \eqref{eq:NR}, we first need to evaluate the second derivative of the log-likelihood with respect to the coefficients $\bm{\alpha}$:
\begin{equation}
\label{eq:hessian}
\frac{\partial^2 \breve{\ell}_P}{\partial \bm{\alpha}^2}=-\bm{B}' \bm{\Gamma} \bm{B}-\bm{P}
\end{equation}

Then we obtain:
\begin{eqnarray*} 
\mbox{LHS} &=& -\bm{B}' \tilde{\bm{\Gamma}} \bm{B} -\bm{P}\\ 
\mbox{RHS}&=&
		-\bm{B}' \tilde{\bm{\Gamma}} \bm{B}
  \tilde{\bm{\alpha}}- \bm{P} \tilde{\bm{\alpha}} 
  - (\bm{B}' \breve{\bm{y}} - \bm{B}' \tilde{\bm{\gamma}} -\bm{P} \tilde{\bm{\alpha}} ) \\
		&=&- \bm{B}' \tilde{\bm{\Gamma}}\tilde{\bm{z}} 
\end{eqnarray*}
where $\tilde{\bm{z}}= \tilde{\bm{\eta}}+ \tilde{\bm{\Gamma}}^{-1}(\tilde{\bm{y}}-\tilde{\bm{\gamma}})$ serves a role analogous to the conventional \emph{working vector} found in the GLM context, albeit evaluated at the scale of the latent response. The final system of equations presented in \eqref{eq:NR} can thus be concisely expressed as follows:
\begin{equation}\label{eq:newIRLS}
(\bm{B}' \tilde{\bm{\Gamma}}  \bm{B}  +\bm{P})  \hat{\bm{\alpha}}  = \bm{B}' \tilde{\bm{\Gamma}}\tilde{\bm{z}}\, .
\end{equation}
This is analogous to the M-step of the EM algorithm, where the penalized log-likelihood of the split data is maximized to obtain $\hat{\bm{\alpha}}$ (and therefore $\hat{\bm{\gamma}})$. 

A key distinction between the estimation approach proposed for the PCLM and the EM algorithm lies in their underlying assumptions about the data distribution. The PCLM approach assumes that the observed aggregated (incomplete) counts, denoted as $\bm{y}$, follow a Poisson distribution with mean $\bm{\mu}$, and estimates an unobservable vector $\bm{\gamma}$, which is related to $\bm{\mu}$ via the composition matrix $\bf{C}$ and is assumed to be smooth.  In contrast, the EM algorithm assumes that the complete (latent) data follow a Poisson distribution with mean $\bm{\gamma}$ and alternates between the redistribution of the counts proportionally
to the current approximation and the estimation of the distribution parameters. 

\subsection{Use of Generalized Linear Array Models in PCLMs}\label{sec:GLAMandCLM}

\citet{CurrieDurbanEilersGLAM2006} introduced an arithmetic of arrays which enables low-storage, high-speed computations within the scoring algorithm of generalized linear models, referred to as the Generalized Linear Array Model (GLAM). In this section, we demonstrate how these principles can be effectively applied to solve \eqref{eq:newIRLS} and estimate coefficients $\bm{\alpha}$. Each iteration of the algorithm requires only two calculations, which can be efficiently performed using the GLAM framework. 

Let's start from the RHS of \eqref{eq:newIRLS}: $\bm{B}'\tilde{\bm{\Gamma}}\tilde{\bm{z}}$. For the calculation of working vector $\bm{z}$, we rewrite in a GLAM setting:
\begin{itemize}
\item The linear predictor:
$$ \tilde{\bm{\eta}}= (\bm{B}_2 \otimes \bm{B}_1) \tilde{\bm{\alpha}} = \verb"vec"(\bm{B}_{1} \tilde{\bm{A}} \bm{B}_{2}') \, ,$$
where $\tilde{\bm{A}}$ is a matrix of size $c_1 \times c_2$ with the elements of $\tilde{\bm{\alpha}}$

\item The latent distribution:
$$\tilde{\bm{\gamma}}=\bm{e} \odot \exp(\tilde{\bm{\eta}})\, ,$$
with $\bm{e}$ being required only when working with mortality data

\item The expected values:
$$\tilde{\bm{\mu}} =\bm{C} \tilde{\bm{\gamma}}= \verb"vec"(\bm{C}_1 \tilde{\bm{\Gamma}}^{*}   \bm{C}_2')\, ,$$
where $\tilde{\bm{\Gamma}}^{*}$ is a matrix of size $m_1 \times m_2$ with the elements of $\tilde{\bm{\gamma}}$.
Note that in this way we avoid constructing the large diagonal matrix $\bm{\Gamma}$. 

\item The working latent response:
$$\tilde{\bm{y}}= \tilde{\bm{\Gamma}}\bm{C}' \tilde{ \bm{W}}^{-1}\bm{y}= (\bm{C}' ( \bm{y}\oslash \tilde{ \bm{\mu}})) \odot \tilde{\bm{\gamma}}= \verb"vec"(\bm{C}_1' \tilde{\bm{M}} \bm{C}_2)\odot \tilde{\bm{\gamma}}\, , $$
where $\tilde{\bm{M}}$ is a $n_1\times n_2$ matrix with the elements of $\bm{y}\oslash \tilde{\bm{\mu}}$.
\end{itemize}
The working vector $\bm{z}$ can now be rewritten as follows:
$$\tilde{\bm{z}}= \tilde{\bm{\eta}}+(\tilde{\bm{y}} -\tilde{\bm{\gamma}})\oslash \tilde{\bm{\gamma}}$$
and consequently the RHS of \eqref{eq:newIRLS} is given by
$$ \bm{B}' \tilde{\bm{\Gamma}} \tilde{\bm{z}}= (\bm{B}_2' \otimes \bm{B}_1') (\tilde{\bm{\gamma}} \odot \tilde{\bm{z}})= \verb"vec"(\bm{B}_1' \tilde{\bm{Z}}\bm{B}_2)\, ,$$
where $\tilde{\bm{Z}}$ is a $m_1\times m_2$ matrix with the elements of $\tilde{\bm{\gamma}} \odot \tilde{\bm{z}}$.

The left-hand side of \eqref{eq:newIRLS}, excluding the penalty term $\bm{P}$, can also be computed efficiently without the need to construct large matrices:
$$\bm{B}' \tilde{\bm{\Gamma}}  \bm{B} \equiv \rho(\mathcal{G}(\bm{B}_2,\bm{B}_2)', \rho(\mathcal{G}(\bm{B}_1,\bm{B}_1),\tilde{\bm{\Gamma^*}})). $$
The symbol $\equiv$ means that both sides have
the same elements but arranged in a different order, the right-hand side of the equation is of size $c_1^2 \times c_2^2$, therefore, it needs to be rearranged into a $c_1c_2\times c_1c_2$ matrix. Furthermore, $\mathcal{G}()$ represents the row-tensor transformation and $\rho()$ denotes the rotated $\mathcal{H}$-transform of an array, both of which were proposed by \citet{CurrieDurbanEilersGLAM2006}. 

The extension to $d$-dimensions is straightforward, with previous calculations adapted as follows:
\begin{itemize}
    \item[a)] Linear functions Linear functions: the elements of $\bm{B} \hat{\bm{\alpha}}$ (and similarly for $\tilde{\bm{\mu}}$, $\tilde{\bm{y}}$  and $\bm{B}' \tilde{\bm{\Gamma}} \tilde{\bm{z}} $ ) are given by the d-dimensional array
$$ \rho(\bm{B}_d, \ldots, \rho (\bm{B}_2, \rho(\bm{B}_1,\tilde{\bm{A}}))\ldots).$$
\item[b)] Inner products: the elements of the inner product  $ \bm{B}' \tilde{\bm{\Gamma}}  \bm{B}$   are given by the d-dimensional array
$$ \rho(\mathcal{G}(\bm{B}_d,\bm{B}_d)', \ldots, \rho (\mathcal{G}(\bm{B}_2,\bm{B}_2)', \rho(\mathcal{G}(\bm{B}_1,\bm{B}_1)',\tilde{\bm{\Gamma^*}}))\ldots).$$
\end{itemize}

A detailed description of these functions, along with \texttt{R} code snippets, is provided in the appendix. Furthermore, a complete set of routines and fully self-reproducible programs for estimating the examples presented in this paper are accessible at \url{osf.io/uwejt/?view_only=2ca1fdb7568342bbb9a3c51fd33c718c}.

\subsection{Uncertainty quantification}\label{sec:uncertainty}

The estimation of the PCLM provides point estimates for the latent mortality rates and associated parameters. However, understanding the variability of these estimates is crucial for accurate interpretation and robust decision-making. This section outlines the approach used for uncertainty quantification, leveraging large-sample results and a Bayesian-inspired framework.

To quantify the uncertainty of the estimate $\hat{\bm{\alpha}}$, we derive its approximate distribution using a Bayesian approach proposed by \cite{Wood2006}. Under this framework:
\begin{equation}\label{eq:SD}
    \bm{\alpha|y} \sim \mathcal{N}\left ( \hat{\bm{\alpha}},  \bm{V}  \right)\,,
\end{equation}
where $\bm{V}=\mathcal{H}^{-1}$ and $\mathcal{H}$ is the observed information matrix (Hessian of the negative penalized log-likelihood)  at $\hat{\bm{\alpha}}$. Some care needs to be taken at this stage since  \eqref{eq:hessian} would correspond to the negative information matrix of the   penalized log-likelihood of $\breve{\bm{y}}$, and so, we would infer that
\begin{equation}
  \bm{\alpha|\breve{y}} \sim \mathcal{N}\left ( \hat{\bm{\alpha}},  (\bm{B}' \bm{\Gamma} \bm{B}+\bm{P})^{-1}\right)\,.   
 \end{equation}

However, we would be underestimating the variance, since we are ignoring the uncertainty due to the redistribution of $\bm{y}$ into $\breve{\bm{y}}$. The correct expression for the information matrix can easily be obtained by calculating the derivative of \eqref{eq:fd1} with respect to $\bm{\alpha}$:
\begin{equation}
\label{hessian2}
 \frac{\partial \, \ell_P}{\partial\, \bm{\alpha}^2}= -\bm{B}' \bm{\Gamma}\bm{C}' \bm{W}^{-1}\bm{C} \bm{\Gamma} \bm{B}'-\bm{P}.   \end{equation}

And so,
\begin{equation}\label{eq:SD2}
    \bm{\alpha|y} \sim \mathcal{N}\left ( \hat{\bm{\alpha}},  (\bm{B}' \bm{\Gamma}\bm{C}' \bm{W}^{-1}\bm{C} \bm{\Gamma} \bm{B}'  +\bm{P})^{-1}\right).
\end{equation}

Furthermore, if the EM algorithm is used, the variance  in \eqref{eq:SD2} corresponds to the expression proposed by \cite{LeePawitanVarianceEM2014}.

In practice, equation \eqref{eq:SD2} is applied by substituting the estimate of $\bm{\gamma}$ (and therefore $\bm{\mu}$) evaluated at $\bm{\alpha}=\hat{\bm{\alpha}}$ once the algorithm in Section \ref{sec:estimation} has reached convergence (in a similar fashion as in the GLM case). Once these substitutions are made, confidence intervals can be calculated. The calculation of the variance-covariance matrix in \eqref{eq:SD2} can be computationally demanding since we can only partially benefit from the GLAM approach, but it is still possible to avoid the computation of matrix $\bm{B}$ (direct calculation of this matrix can be intractable when the dimension of the array increases). 

In general, our main objective is to compute confidence intervals for $\hat{\bm{\eta}}=\bm{B}\hat{\bm{\alpha}}$. To accomplish this, it suffices to extract the diagonal elements of $\bm{B}\bm{V}\bm{B}'$, which requires the following steps:
\begin{itemize}
    \item Calculation of $\bm{V}$:
  $$\bm{V}=(\bm{B}' \bm{\Gamma}\bm{C}' \bm{W}^{-1} \bm{C} \bm{\Gamma} \bm{B}  +\bm{P})^{-1}.$$ 
This implies computing the inner product $\bm{B}' \bm{\Gamma}\bm{C}' \bm{W}^{-1} \bm{C} \bm{\Gamma} \bm{B}$, which unfortunately can't be rewritten in a GLAM setting, but we can still avoid calculating $\bm{B}$ since:
$$\bm{B}' \bm{\Gamma}\bm{C}'\equiv (\bm{B}_2' \otimes \bm{B}_1')\bm{\Gamma}(\bm{C}_2' \otimes \bm{C}_1') \equiv \rho(\mathcal{G}( \bm{B}_2, \bm{C}_2')', \rho(\mathcal{G}(\bm{B}_1,\bm{C}_1'),\bm{\Gamma}^*)), $$
where $\bm{\Gamma}^{*}$ is a matrix of size $m_1 \times m_2$ with the elements of $\hat{\bm{\gamma}}$. In this case, the matrix obtained by using GLAM methods is of size $c_1n_1\times c_2n_2$ and needs to be reshaped into a matrix of size $c_1c_2\times n_1n_2$. To complete the computation of $\bm{V}$, we calculate the product $\bm{C}\bm{\Gamma}\bm{B}$, which is simply the transpose of the previously obtained $\bm{B}'\bm{\Gamma}\bm{C}'$. The final inner product with $\bm{W}^{-1}$ can be performed efficiently without fully constructing $\bm{W}$, as it is diagonal.

\item Calculation of diagonal elements of $\bm{B}\bm{V} \bm{B}'$
 $$\verb"diag"(\bm{B}\bm{V} \bm{B}')= \verb"diag"((\bm{B}_1' \otimes \bm{B}_2')\bm{V}(\bm{B}_1 \otimes \bm{B}_2))= \verb"vec"(\rho(\mathcal{G}(\bm{B}_1)', \rho(\mathcal{G}(\bm{B}_2),\bm{V^*}))), $$
where $\bm{V^*}$ is the result of reorganizing the elements of $\bm{V}$ into a $c_1^2 \times c_2^2$ matrix.
\end{itemize}

A notable computational advantage of our approach is that the matrix $\bm{V}$ is computed only once, after the algorithm has converged. In contrast, the original estimation procedure proposed by \cite{ThompsonBakerCLM1981} and later adopted by \cite{Eilers2007} requires $\bm{V}$ to be recalculated at every iteration of the algorithm, adding considerable computational overhead.

\section{Simulation Studies}\label{sec:Simulations}

We conduct two simulation studies designed to evaluate the performance of the proposed PCLM methodology. The simulation setup is closely aligned with our motivating applications and aims to reflect realistic grouped data scenarios. Specifically, we generate high-resolution latent rates over a regular grid, apply a predefined aggregation scheme to produce grouped intensities, and simulate observed counts from a Poisson distribution. The following diagram shows the data generation process:

\begin{center}
   \begin{tikzpicture}[
  node distance=1.8cm and 1.2cm,
  every node/.style={draw, minimum height=1.2cm, minimum width=2cm, align=center},
  every path/.style={->, thick, >=Stealth}
]

\node (eta) {Design \\$\bm{\gamma}=\exp(\bm{\eta})$};
\node (gamma) [right=of eta] {Aggregate \\$\bm{\mu} = \bm{C\gamma}$};
\node (y) [right=of gamma] {Simulate \\$\bm{y} \sim \mathcal{P}(\bm{\mu})$};

\draw (eta) -- (gamma);
\draw (gamma) -- (y);

\end{tikzpicture}

\end{center}

In the first subsection, we replicate this process multiple times within a two-dimensional setting under various underlying scenarios to evaluate the performance of the PCLM. In Section~\ref{sec:CompareComput}, we extend the analysis to three- and four-dimensional settings to demonstrate the computational advantages of the proposed approach. 

\subsection{Assessing the performance of the Penalized Composite Link Model}\label{sec:ComparePCLM}

In this section, we evaluate the performance of the PCLM in ungrouping Poisson-distributed data within a two-dimensional framework, where observations are aggregated along both dimensions at varying levels of grouping that mimic real-world data applications. Similar evaluations have been conducted in previous studies for one-dimensional \citep{RizziEtAlPCLM1D2015, LambertEilersBayesGrouped2009} and two-dimensional \citep{LambertBayesBivariateCSDA2011} settings.

To conduct a comprehensive assessment under controlled yet realistic conditions, we devised two simulation scenarios, labeled \textbf{A} and \textbf{B}, each defined by a distinct true underlying linear predictor $\bm{\eta}$. In both scenarios, data are simulated over a two-dimensional covariate grid defined by $\bm{x}_{1} = 1,\dots,m_1=80$ and $\bm{x}_{2} = 1,\dots,m_2=60$, resulting in a total of $m_1 \times m_2 = 4{,}800$ latent observations.

Scenario A uses a synthetic specification of $\bm{\eta}$ designed to reflect plausible nonlinear and interaction structures. Specifically, we define two coefficient functions over the temporal dimension $\bm{x}_2$: $\bm{\beta}_1 = -10 + 0.5\cos(\bm{x}_2 / 40), \quad \bm{\beta}_2 = 0.1 + 0.025\cos(\bm{x}_2 / 40)$. And we set a fixed vector $\bm{\beta}_3 = \bm{1}_{n_2}$. The covariate information is represented by the design matrix $\bm{X} = \left[\bm{1}_{m_1} : \bm{x}_1 : -\sin\left(\frac{\pi \bm{x}_1}{50}\right)\right]$. This setting corresponds to a varying-coefficient model in which both the intercept and slope vary smoothly with $\bm{x}_{2}$. It is designed to emulate real-world mortality patterns, where rates change gradually and non-linearly across age and time. The structure allows for a controlled and interpretable representation, with linear predictor values ranging approximately from $-10$ to 1.5.

Scenario B, in contrast, relies on empirically observed demographic data: we use age-specific mortality rates for Swedish females from age 20 to 99 ($\bm{x}_1$) and years 1960--2019 ($\bm{x}_2$), applying light $P$-spline smoothing to obtain a realistic underlying linear predictor. The dimensional structure in this scenario matches that of Scenario A, with $\bm{x}_1$ and $\bm{x}_2$ forming an $80 \times 60$ grid. In this case, the smoothed linear predictor exhibits values ranging from $-8.4$ to $-0.7$, reflecting typical mortality patterns on the log scale.

In both cases, the predictor $\bm{\eta}$ is used to compute a vector of latent distribution $\bm{\gamma}$, and expected Poisson counts are obtained by incorporating an exposure matrix $\bm{e}$, which varies across both covariate dimensions. Although exposures simply act as an offset in the model, we design $\bm{e}$ to emulate real populations, with decreasing exposure by $\bm{x}_1$ and smoothly varying patterns over $\bm{x}_2$. Two exposure settings (\textbf{small} and \textbf{large}) are considered, differing by a factor of 20 in magnitude. To contextualize the large exposure setting, it corresponds to population sizes ranging approximately from 15 to 30 million. This setup yields four distinct configurations, defined by the combination of two linear predictor scenarios (A and B) with the two exposure levels (small and large).

To assess the robustness of the PCLM under data aggregation, we simulate counts from the latent Poisson distribution and apply grouping along both dimensions using interval widths of 1, 2, 5, and 10, resulting in 16 distinct grouping schemes, including the fully ungrouped case. For illustration, the coarsest grouping---using interval widths of 10 along both dimensions---produces only $n_1 = 8$ and $n_2 = 6$ observed cells, totaling 48 aggregated data points. In contrast, grouping with interval widths of 5 yields 192 observed values ($n_1 = 16$, $n_2 = 12$). Across all configurations, the goal remains the same: to recover the original latent distribution defined over the full grid of size $m_1 \times m_2 = 4{,}800$.

For each of the 64 (4~$\times$~16) setting combinations, we apply the PCLM, use rich $B$-spline bases over the two dimensions ($c_1=16$ and $c_2=12$) and optimize the smoothing parameters via minimization of the Bayesian Information Criterion \citep{Schwarz1978}. This enables estimation of the underlying $\bm{\eta}$ across various grouping structures. Each configuration is replicated 100 times, and we compute the Root Mean Square Error (RMSE) between the estimated and true $\bm{\eta}$ to assess reconstruction accuracy. 

\begin{figure}
\centering
\includegraphics[scale=0.4]{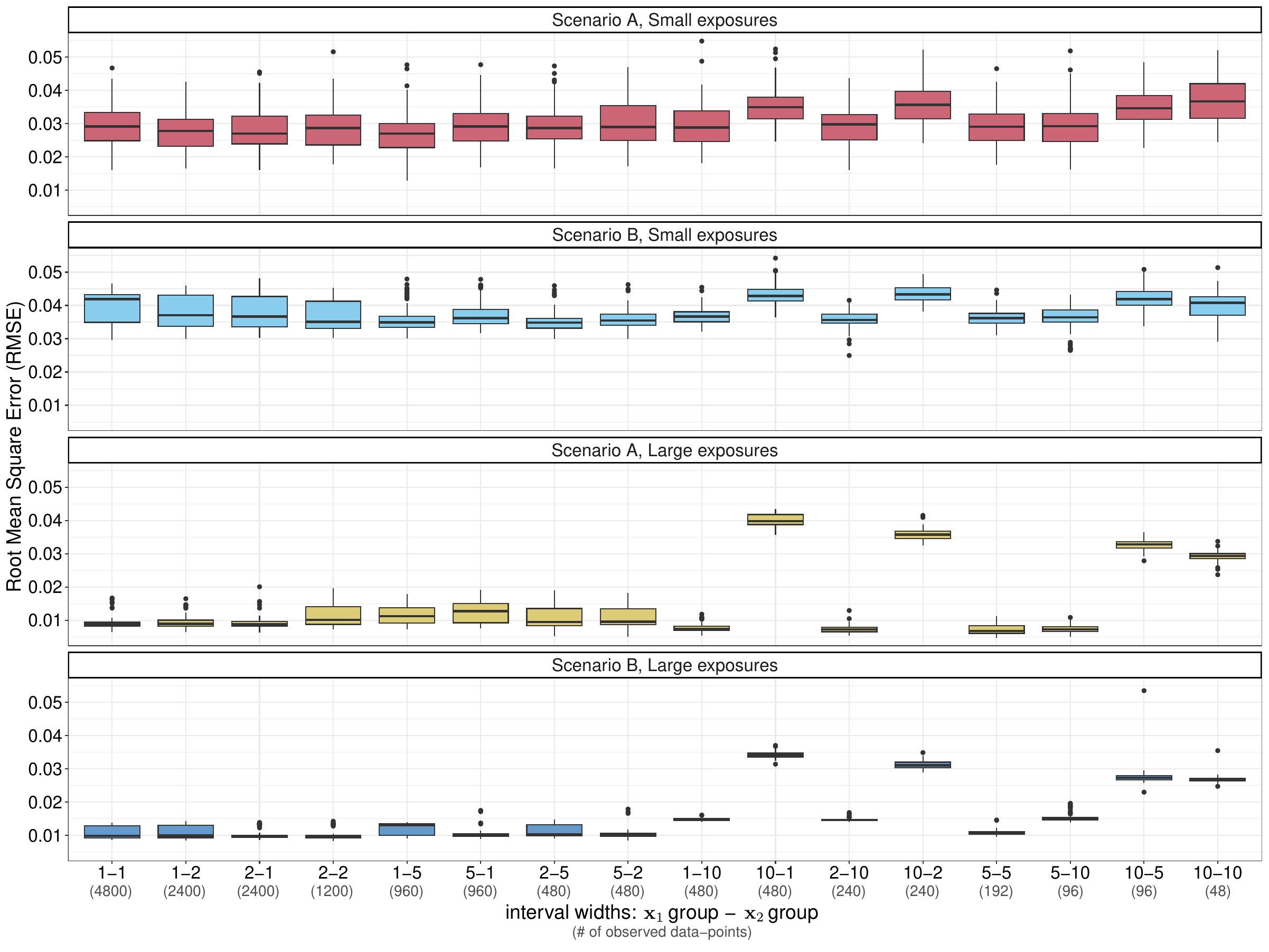}
\caption{Root Mean Square Error (RMSE) of the estimated linear predictor relative to the true underlying values, based on 100 simulation replicates for each setting in a two-dimensional framework. Boxplots show the distribution of RMSE across different grouping schemes, defined by combinations of interval widths along the two dimensions, which correspond to different numbers of observed data points. Each panel represents one of the four combinations of linear predictor scenarios (A and B) and exposure levels (Small and Large).}
\label{fig:RMSEsimulation}
\end{figure}

Outcomes are shown in Figure~\ref{fig:RMSEsimulation} across the 16 different grouping schemes and four simulation settings. Each boxplot summarizes the distribution of RMSE over 100 replicates for a given grouping configuration, denoted by the interval widths over the two dimensions $\bm{x}_1$ and $\bm{x}_2$. Overall, across all simulation settings, RMSE values remain small, even under substantial data aggregation, indicating that the PCLM performs effectively in ungrouping and smoothing Poisson-distributed data observed in grouped form. Numerically, RMSE values range from approximately 0.01 to 0.05. Given that the true linear predictor in both scenarios spans a range from $-10$ to 1.5, these errors correspond to less than 0.5\% of the total scale.

However, a clear, albeit more modest than anticipated, trend emerges: as the level of aggregation increases (e.g., interval widths of 5 or 10), estimation error tends to rise. However, even with coarser groupings, the resulting RMSE values remain moderate, indicating that the PCLM retains robustness under substantial aggregation. In contrast, finer groupings (e.g., 1--2, 2--1) consistently yield lower RMSE values, particularly under the large exposure setting, approaching the accuracy observed with fully disaggregated data (1--1). The influence of exposure level is more pronounced: for both scenarios, RMSE values are systematically lower and exhibit reduced variability when exposure is large, reflecting the stabilizing effect of higher expected counts on estimation accuracy.

Scenario A, which is based on analytically constructed data, generally yields slightly lower RMSE values compared to Scenario B. This likely reflects the more regular and controlled structure of the underlying linear predictor in Scenario A. Overall, these results confirm that the PCLM is robust in recovering the true linear predictor, even under substantial data aggregation, particularly when exposure levels are high and grouping intervals are not excessively coarse.

\subsection{Comparing computational performances}\label{sec:CompareComput}

To assess the computational efficiency of our proposed GLAM-based implementation of the PCLM, we conducted a dedicated simulation study focusing on differences in storage and runtime performance between the original algorithm and the GLAM-based alternative. In contrast to previous simulations, where the emphasis was on estimation accuracy, the goal here is to evaluate how each implementation scales computationally when applied to higher-dimensional data structures.

We considered data structures of increasing dimensionality, ranging from two to four dimensions, each defined over regularly spaced covariate grids. To reflect more realistic scenarios, we examined a relatively large data configuration where $\bm{x}_1 = \bm{x}_2 = 1, \dots, 50$ and $\bm{x}_3 = \bm{x}_4 = 1, \dots, 20$, i.e., $m_1 = m_2 = 50$ and $m_3 = m_4 = 20$, resulting in a latent vector $\bm{\gamma}$ of length 2,500, 50,000, and 1,000,000 in the two-, three-, and four-dimensional cases, respectively. For comparison purposes, we also considered a smaller, simplified example with $m_1 = m_2 = 40$ and $m_3 = m_4 = 8$, for which the corresponding lengths of $\bm{\gamma}$ are 1,600, 12,800, and 102,400.

Over each of these domains and both data structures, we constructed smooth univariate functions as follows: $\bm{f}_1 = (\sin(\bm{x}_1/20) + 1)/2$, $\bm{f}_2 = -4(\cos(\bm{x}_2/20) + 1)$, $\bm{f}_3 = \sin(\bm{x}_3/30)$, and $\bm{f}_4 = \cos(\bm{x}_4/40)$. The true underlying linear predictor was constructed in each setting by applying successive outer products across dimensions, starting from the two-dimensional case (using $\bm{f}_1$ and $\bm{f}_2$) and progressively incorporating $\bm{f}_3$, and $\bm{f}_4$ for the three-, and four-dimensional settings, respectively.

From the resulting linear predictor arrays, Poisson-distributed count data were simulated. To maintain comparability with earlier simulations and future applications, expected values for the Poisson distribution were defined using both the exponential of the linear predictor and an exposure term. In this setting, exposures were kept constant at $10{,}000$ across the entire covariate grid as a simplifying assumption. For all configurations, we applied grouping only along the first two dimensions ($\bm{x}_1$ and $\bm{x}_2$) by aggregating every five consecutive units, thereby introducing coarsening that mirrors real-world data aggregation practices. Specifically, the resulting dimensions are $n_1=n_2=12$ in the realistic scenario, and $n_1=n_2=8$ in the smaller scenario.

To isolate differences in computational performance, we fixed the model complexity across all dimensions by selecting a relatively rich tensor-product $B$-spline basis with internal knots every 5 available data-points along each corresponding dimension. For simplicity and consistency, smoothing parameters were fixed at a subjectively chosen value of 100 across all dimensions. This setting allows a fair and focused comparison of memory usage and computation time between the standard and GLAM-based implementations of the PCLM across varying dimensional complexities. While computer performance was not explicitly tested in previous simulation studies, all models for this study, as well as for the applications presented in Section~\ref{sec:Applications}, were run on a portable personal computer equipped with an Intel i7-10610U processor (1.8 GHz) and 16 GB of RAM. This detail is important for interpreting the observed memory demands and computational times.

\begin{table}[h!]
\begin{center}
\begin{tabular}{ccl|rr|rr}
\multicolumn{1}{l}{} &
& \multicolumn{1}{c|}{}           
& \multicolumn{2}{c|}{Memory usage (MB)} 
& \multicolumn{2}{c}{Computation time (s)}\\ 
\multicolumn{1}{c|}{Data}
& \multicolumn{1}{c|}{\begin{tabular}[c]{@{}c@{}}Dimension \\ (length of $\bm{\gamma}$)\end{tabular}} 
& \multicolumn{1}{c|}{Algorithm:} 
& \multicolumn{1}{c}{\begin{tabular}[c]{@{}c@{}}All \\ objects\end{tabular}} 
& \multicolumn{1}{c|}{\begin{tabular}[c]{@{}c@{}}w/o variance \\ estimation\end{tabular}} 
& \multicolumn{1}{c}{\begin{tabular}[c]{@{}c@{}}Complete \\ estimation\end{tabular}} 
& \multicolumn{1}{c}{\begin{tabular}[c]{@{}c@{}}w/o variance \\ estimation\end{tabular}} \\ \hline
\multicolumn{1}{c|}{\multirow{6}{*}{Small}} 
& \multicolumn{1}{c|}{\multirow{2}{*}{\begin{tabular}[c]{@{}c@{}}2D \\ (1,600)\end{tabular}}} 
& \cellcolor[HTML]{EFEFEF}GLAM-based
& \cellcolor[HTML]{EFEFEF} 0.57
& \cellcolor[HTML]{EFEFEF} 0.43
& \cellcolor[HTML]{EFEFEF} 0.05
& \cellcolor[HTML]{EFEFEF} 0.04
\\
\multicolumn{1}{c|}{} & \multicolumn{1}{c|}{}      
& Original
& 22.47                       
& 1.95                        
& 0.26                      
& 0.05
\\ \cline{2-7} 
\multicolumn{1}{c|}{} & \multicolumn{1}{c|}{\multirow{2}{*}{\begin{tabular}[c]{@{}c@{}}3D \\ (12,800)\end{tabular}}} 
& \cellcolor[HTML]{EFEFEF}GLAM-based                      
& \cellcolor[HTML]{EFEFEF} 7.30                           
& \cellcolor[HTML]{EFEFEF} 4.05                          
& \cellcolor[HTML]{EFEFEF} 0.36                             
& \cellcolor[HTML]{EFEFEF} 0.30
\\
\multicolumn{1}{c|}{}& \multicolumn{1}{c|}{}           
& Original    
& 1394.76
& 82.89
& 43.77
& 7.16
\\ \cline{2-7} 
\multicolumn{1}{c|}{} & \multicolumn{1}{c|}{\multirow{2}{*}{\begin{tabular}[c]{@{}c@{}}4D \\ (102,400)\end{tabular}}} 
& \cellcolor[HTML]{EFEFEF}GLAM-based
& \cellcolor[HTML]{EFEFEF} 148.78                     
& \cellcolor[HTML]{EFEFEF} 64.07                  
& \cellcolor[HTML]{EFEFEF} 13.81                  
& \cellcolor[HTML]{EFEFEF} 9.21                  
\\
\multicolumn{1}{c|}{} & \multicolumn{1}{c|}{}          
& Original                      
& {\scriptsize \emph{OOM}}                              
& 4290.90        
& \xmark
& 1894.46
\\ \hline
\multicolumn{1}{c|}{\multirow{6}{*}{Large}} & \multicolumn{1}{c|}{\multirow{2}{*}{\begin{tabular}[c]{@{}c@{}}2D \\ (2,500)\end{tabular}}} 
& \cellcolor[HTML]{EFEFEF}GLAM-based                      
& \cellcolor[HTML]{EFEFEF} 1.93                             
& \cellcolor[HTML]{EFEFEF} 1.24
& \cellcolor[HTML]{EFEFEF} 0.16
& \cellcolor[HTML]{EFEFEF} 0.14
\\
\multicolumn{1}{c|}{}& \multicolumn{1}{c|}{}   
& Original
& 113.39                     
& 9.52
& 1.55
& 0.35
\\ \cline{2-7} 
\multicolumn{1}{c|}{} & \multicolumn{1}{c|}{\multirow{2}{*}{\begin{tabular}[c]{@{}c@{}}3D \\ (50,000)\end{tabular}}} 
& \cellcolor[HTML]{EFEFEF}GLAM-based                      
& \cellcolor[HTML]{EFEFEF} 52.69                           
& \cellcolor[HTML]{EFEFEF} 20.26
& \cellcolor[HTML]{EFEFEF} 2.31
& \cellcolor[HTML]{EFEFEF} 1.23
\\
\multicolumn{1}{c|}{} & \multicolumn{1}{c|}{}          
&  Original                   
& {\scriptsize \emph{OOM}}        
& 2022.93       
& \xmark
& 483.19
\\ \cline{2-7} 
\multicolumn{1}{c|}{} & \multicolumn{1}{c|}{\multirow{2}{*}{\begin{tabular}[c]{@{}c@{}}4D \\ (1,000,000)\end{tabular}}} 
& \cellcolor[HTML]{EFEFEF}GLAM-based                      
& \cellcolor[HTML]{EFEFEF} 2588.91
& \cellcolor[HTML]{EFEFEF} 369.09
& \cellcolor[HTML]{EFEFEF} 525.35
& \cellcolor[HTML]{EFEFEF} 260.44
\\
\multicolumn{1}{c|}{} & \multicolumn{1}{c|}{}          
& Original
& {\scriptsize \emph{OOM}}
& {\scriptsize \emph{OOM}}
& \xmark
& \xmark  
\end{tabular}
\caption{Memory usage (MB) and computation time (s) for estimating a Penalized Composite Link Model across two-, three-, and four-dimensional data structures. Results compare the proposed GLAM-based algorithm with the original estimation procedure. The label {\scriptsize \emph{OOM}} indicates an Out Of Memory error, meaning the procedure failed due to insufficient memory on the test machine. The symbol \xmark\ denotes that the corresponding value could not be computed for this reason.}\label{tab:ComputationSimul}
\end{center}
\end{table}

The results in Table~\ref{tab:ComputationSimul} clearly highlight the computational advantages of the proposed GLAM-based algorithm over the original estimation procedure for Penalized Composite Link Models (PCLM) across varying data dimensions and sizes. In all scenarios considered, the GLAM-based approach consistently demonstrates substantially lower memory usage and faster computation times. These improvements become more pronounced as the dimensionality and data size increase. For instance, in the small 3D case ($|\bm{\gamma}| = 12{,}800$), the original algorithm requires approximately 1{,}395~MB of memory for full estimation, whereas the GLAM-based algorithm completes the same task using just 7.3~MB, yielding a nearly 190-fold reduction. Likewise, computation time drops from 43.77 seconds to 0.36 seconds.

For relatively large settings, while the original algorithm is sometimes able to estimate the parameter vector $\bm{\gamma}$, it often fails to compute the associated variance-covariance matrix due to excessive memory requirements. This limitation prevents quantification of the uncertainty in the estimates, which is critical for inference and practical applications. For example, in the small 4D case ($|\bm{\gamma}| = 102{,}400$), the GLAM-based method completes full estimation—including variance estimation—in approximately 13.81 seconds, whereas the original approach requires over 31 minutes (1894.46 seconds) just to estimate the point estimates without successfully computing the variance-covariance matrix.

The benefits of the GLAM-based approach are even more apparent in higher-dimensional scenarios. In the large 4D setting ($|\bm{\gamma}| = 1{,}000{,}000$), the original algorithm is unable to proceed at all, whereas the proposed method, while demanding in absolute terms (around 2.6~GB and 525 seconds), remains fully operational and scalable.

Overall, the GLAM-based algorithm markedly improves the feasibility of fitting PCLMs in high-dimensional settings. It allows users to perform comprehensive estimation tasks on large and complex datasets that would otherwise exceed memory limitations with the original method. 

In terms of memory usage, the primary bottleneck of the original algorithm lies in the construction of the full model and composite matrices. For example, in a relatively small four-dimensional scenario, the model matrix 
$\bm{B} = \bm{B}_4 \otimes \bm{B}_3 \otimes \bm{B}_2 \otimes \bm{B}_1$ and the composite matrix $\bm{C} = \mathrm{diag}(m_4) \otimes \mathrm{diag}(m_3) \otimes \bm{C}_2 \otimes \bm{C}_1$ require approximately 800~MB and 3,200~MB of memory, respectively. Consequently, even before any computation, the storage of $\bm{B}$ and $\bm{C}$ represents a significant constraint, especially in resource-limited environments such as personal computers.

These results highlight not only the efficiency but also the robustness of the proposed method for practical applications involving multi-dimensional smoothing and large-scale data where count data are grouped, and the objective is to estimate latent underlying distributions.

\section{Applications}\label{sec:Applications}

\subsection{Mortality grouped by age and years}

In this section, we present an illustrative example primarily aimed at demonstrating the computational efficiency of the proposed algorithm for estimating a PCLM, which has been reformulated to leverage GLAM arithmetic. We construct an example where the underlying distribution over single years of age and year is known. We then artificially aggregate the observed counts to simulate grouped data, allowing us to focus on computational aspects.

We use ungrouped mortality data for Swedish females obtained from the \citet[HMD,][]{HMD}, spanning the years 1960 to 2019 and ages 10 to 104. The data are originally reported at a fine resolution, with death counts and population exposures recorded by single year of age and single calendar year. This structure defines a high-resolution grid of size \( m_1 = 95 \) (ages) by \( m_2 = 60 \) (years), yielding a total of 5{,}700 latent data points, which are later smoothed using a standard two-dimensional $P$-spline approach for illustrative purposes.

\begin{figure}
\centering
\includegraphics[scale=0.32]{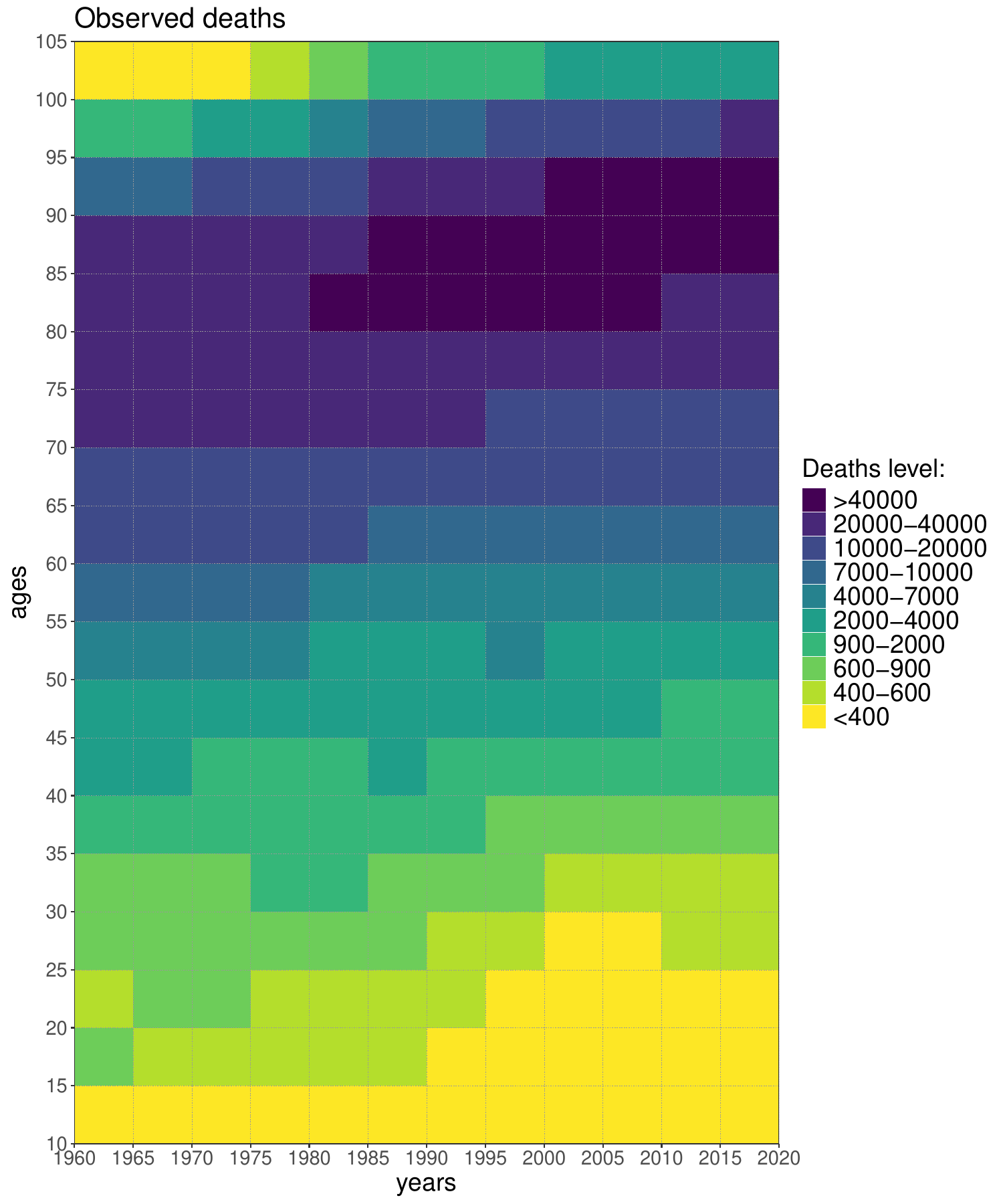}
\includegraphics[scale=0.32]{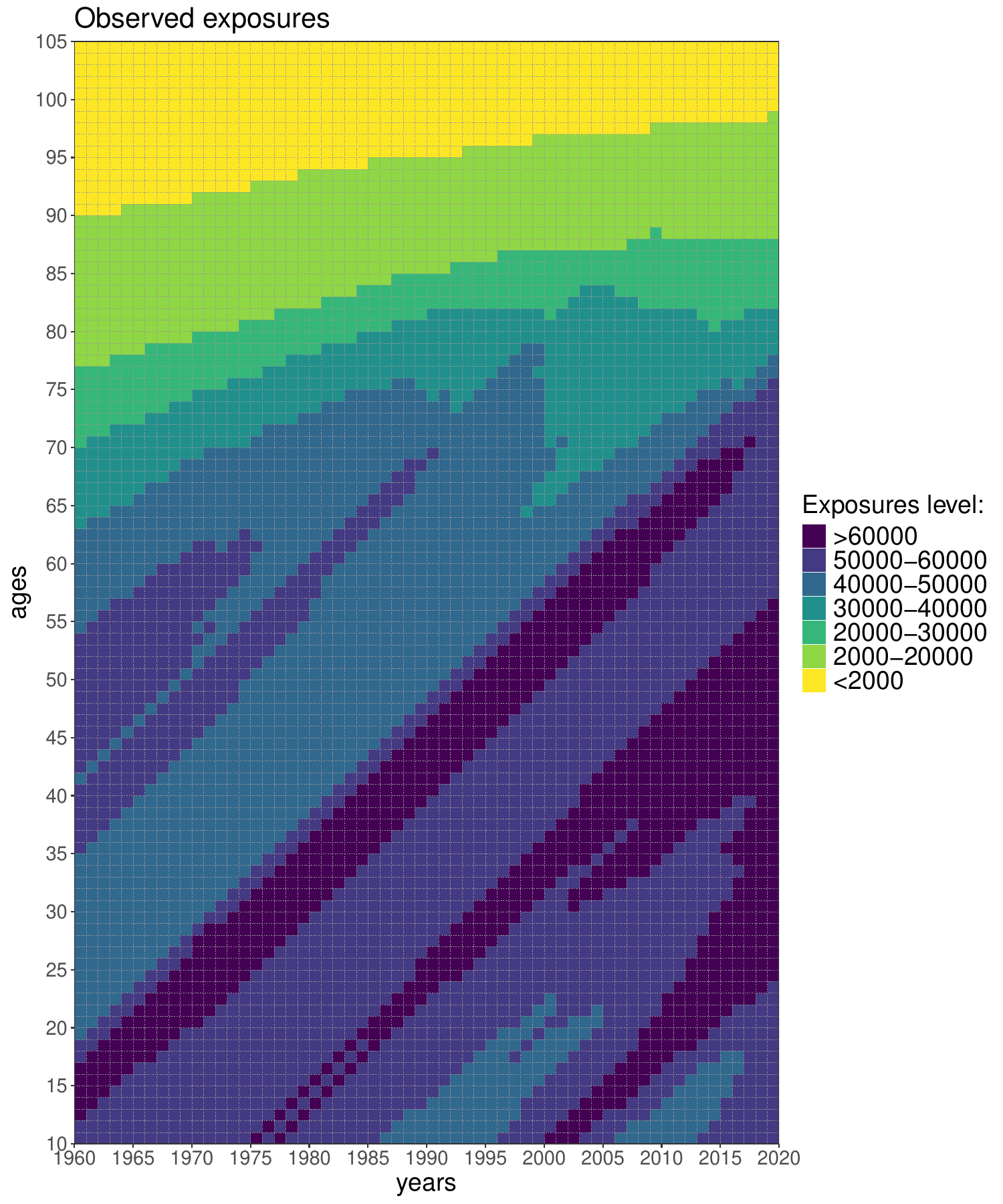}
\caption{Left panel: Observed deaths, aggregated into 5-year age groups and 5-year calendar intervals. Right panel: Observed exposures by single year of age and calendar year. Data for females in Sweden, ages 10–104, from 1960 to 2019. Death counts originally provided by single year of age and calendar year have been aggregated for illustrative purposes.}
\label{fig:ActualDeathsExposures}
\end{figure}

For our analysis, we aggregate the original death counts into 5-year age groups and 5-year calendar intervals. This results in an observed dataset with \( n_1 = 19 \) age groups and \( n_2 = 12 \) time intervals, for a total of \( n_1 \times n_2 = 228 \) observed cells:
$$
\bm{Y} = \begin{bmatrix}
 y_{10-14, 1960-64} & y_{10-14,1965-69} & \cdots & y_{10-14, 2015-19} \\
 y_{15-19, 1960-64} & y_{15-19, 1965-69} & \cdots & y_{15-19, 2015-19} \\
 \vdots & \vdots & \ddots & \vdots \\
 y_{100-104, 1960-64} & y_{100-104, 1965-69} & \cdots & y_{100-104, 2015-19} \\
 \end{bmatrix} \, .
$$
Note that the population exposures were retained at their original fine scale and used as offsets in the model, consistent with the common availability of detailed population denominators even when mortality data are aggregated. Figure~\ref{fig:ActualDeathsExposures} displays the actual death counts (left panel) and population exposures (right panel) as shaded contour plots across age and time, i.e.~Lexis surfaces. The difference in the level of aggregation between the two datasets is clearly noticeable. 

\begin{figure}
\centering
\includegraphics[scale=0.35]{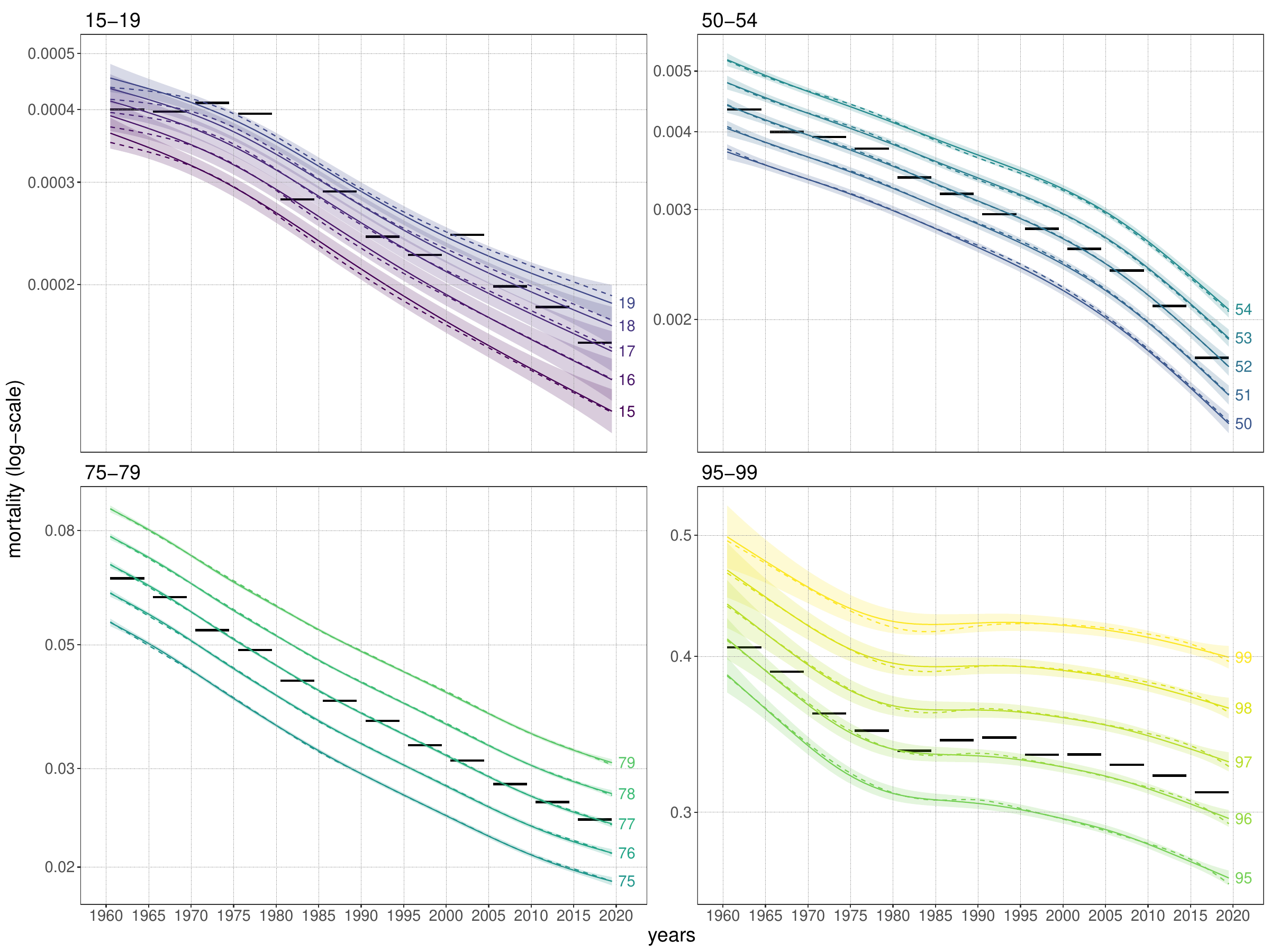}
\caption{Observed death rates by 5-year age groups and 5-year calendar intervals, along with estimated ungrouped mortality by single year of age and calendar year, for four selected age groups (15–19, 50–54, 75–79, and 95–99) over time with associated 95\% confidence intervals. Data for females in Sweden, ages 10–104, from 1960 to 2019. For illustrative purposes, we also include (in dashed lines) the estimates that would have been obtained using a standard two-dimensional $P$-splines approach applied directly to the original unaggregated data.}
\label{fig:ActualSmoothRatesYears}
\end{figure}

\begin{figure}
\centering
\includegraphics[scale=0.5]{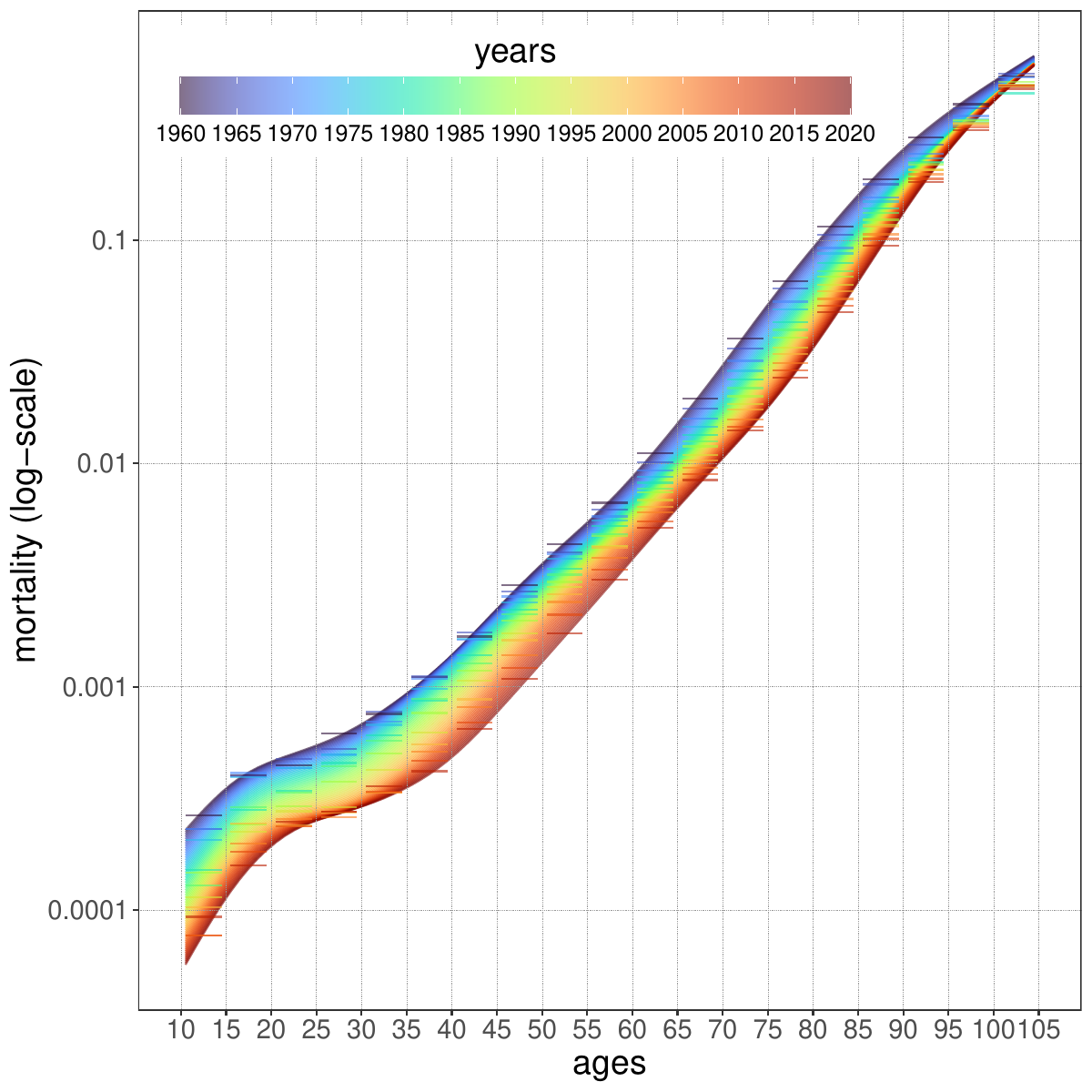}
\caption{Observed death rates by 5-year age groups and 5-year calendar intervals, along with estimated ungrouped mortality by single year of age and calendar year over age. Data for females in Sweden, ages 10–104, from 1960 to 2019.}
\label{fig:ActualSmoothRatesAges}
\end{figure}

By applying a PCLM to these data, we estimate the latent mortality, $\bm{\gamma}$, for each $m_1=95$ single year of age and $m_2=60$ calendar year. Figure~\ref{fig:ActualSmoothRatesYears} illustrates selected outcomes of this approach, displaying log-mortality over time for four chosen age groups. Unlike the Lexis surfaces, this representation reveals both the uncertainty associated with each estimated time trend (95\% confidence intervals) and the disaggregation capability of the PCLM: while observed log-rates are grouped into five-by-five age-year categories, our estimates provide continuous values for each single year and age. Additionally, note that to visualize observed log-rates, exposures must be aggregated to match the death data grouping level, resulting in some loss of information. Similarly Figure~\ref{fig:ActualSmoothRatesAges} presents observed and estimated mortality over ages. Here it is clear how the model is able to describe mortality age-patterns for all available years.  

To obtain our results, we used $c_1=19$ $B$-splines over the age domain and $c_2=12$ $B$-splines over the time domain. The penalty term in \eqref{eq:Pterm} was constructed using second-order differences, with smoothing parameters set to $(\lambda_1, \lambda_2)=(10, 1000)$. Under these settings, the effective dimension of the estimated model was determined to be 65, derived from an initial dataset of $n_1n_2=228$ data points and an estimated latent vector $\bm{\gamma}$ of length $m_1m_2=5700$.

For illustrative purposes, we additionally apply a standard two-dimensional $P$-spline smoothing technique directly to the original ungrouped mortality data. The resulting estimates are shown as dashed lines in Figure~\ref{fig:ActualSmoothRatesYears}. This smoothed surface, obtained from the full-resolution data (5,700 data points), serves as a reference benchmark. It enables a visual and quantitative comparison to the estimates produced by the proposed PCLM when fitted to substantially more coarsely grouped data (228 aggregated observations). This comparison helps assess the model's ability to recover the essential structure and temporal evolution of the underlying mortality surface, even in the presence of considerable information loss due to aggregation.

To estimate the PCLM on the coarsely grouped data, we implement both the original iterative procedure introduced by \citet{ThompsonBakerCLM1981} and later extended by \citet{Eilers2007}, as well as the computationally efficient approach proposed in this paper. While both methods produce equivalent results, thereby confirming the correctness of our implementation, the proposed GLAM algorithm offers a significant advantage. Its primary strength lies in its exceptional computational efficiency, making it a superior choice for large-scale applications. In the previous two-dimensional illustrative dataset, fitting the model with the proposed method and described settings takes approximately 0.14 seconds, compared to 7.9 seconds for the original algorithm. Notably, computing the variance-covariance matrices accounts for about 24\% and 79\% of the overall computational time for the proposed and original approach, respectively. Excluding this step, the time required for model fitting alone is reduced to about 1.7 seconds for the original method and just 0.11 seconds for the proposed approach. Furthermore, the proposed approach is highly efficient in terms of data storage. With the proposed method, the combined size of all relevant \texttt{R} objects used for data and estimation is only 4.4 MB, compared to 284 MB required by the conventional algorithm, a substantial reduction in memory usage.

Although these improvements may seem modest given current computational power, the proposed approach offers a clear resolution to potential storage limitations and delivers remarkable speed enhancements, achieving gains by orders of magnitude over direct evaluation. Furthermore, in practical applications, the search for optimal smoothing parameters often involves numerous repetitions of the scoring algorithm, where the substantial reduction in computational time becomes increasingly impactful. 

\subsection{Mortality grouped by age over years and weeks}

For this application, we present an actual dataset for Spanish males in which mortality is analyzed by age, year, and week. Specifically, data on deaths are available by age group (with intervals of 5 years: 0-4, 5-9, $\ldots$, 85-89, 90+) and across the years 2000 to 2019, as well as by week (1-52). These death counts were sourced from the \cite{INEweekData} and provided by the \citet{STMF2024}.

The dataset for the exposures was obtained from the \citet{HMD}, where data are available by single years of age, ranging from 0 to 104 years, and for each year. To ensure the temporal consistency of the dataset for analysis, death counts were adjusted to align with a 52-week year, and the original annual exposure data were linearly interpolated to achieve a weekly resolution.

The resulting dataset is three-dimensional, with grouping observed only over the age dimension. Specifically, the dimensions of the dataset and the model are $n_1=19$ and $m_1=105$, corresponding to the number of age groups and the maximum age of 104, derived from the exposures' available ages. The dimensions for the years and weeks are $n_2=m_2=20$ and $n_3=m_3=52$, respectively. This structured dataset allows for a comprehensive modeling of mortality patterns across different age groups, years, and weeks.

\begin{figure}
\centering
\includegraphics[scale=0.5]{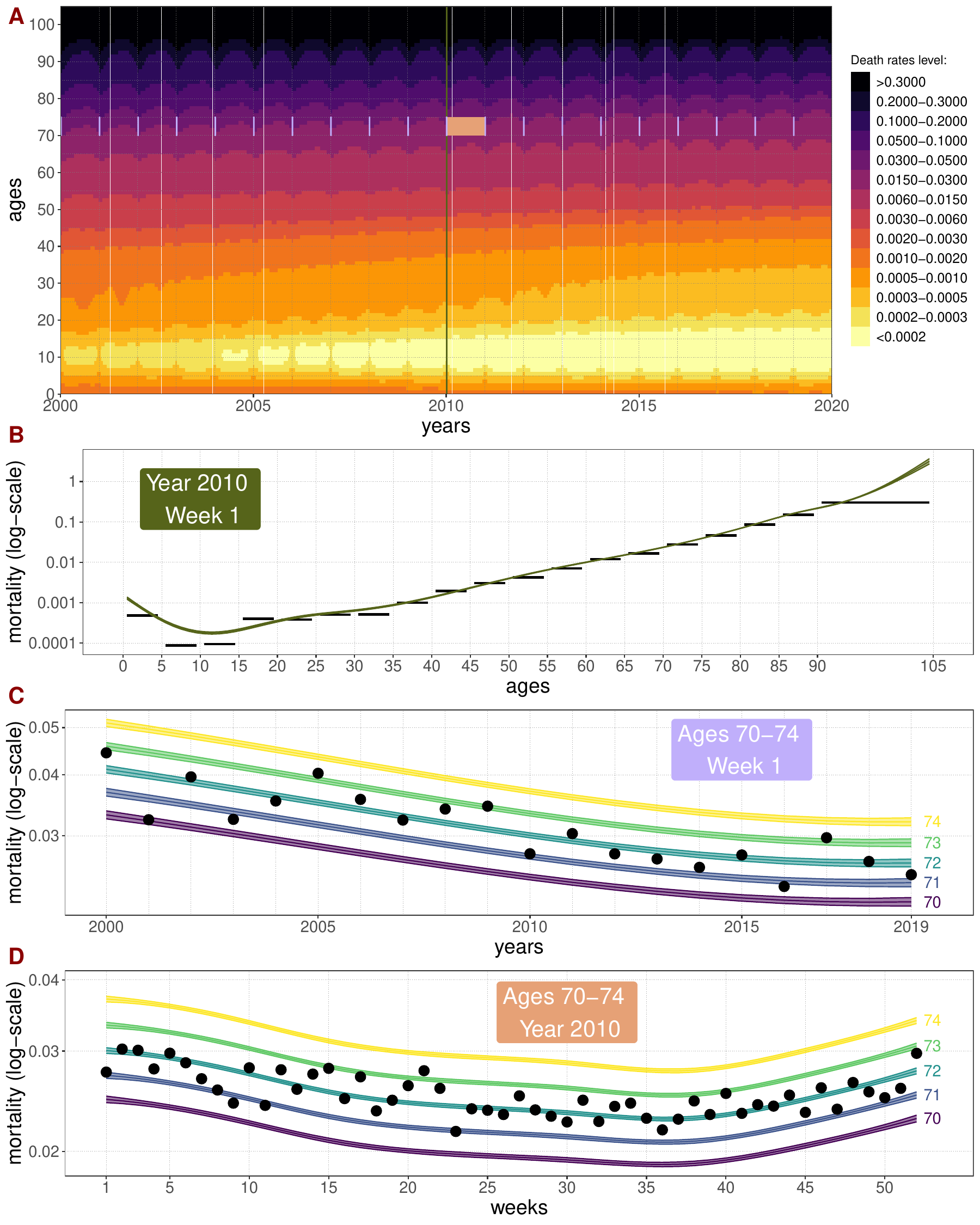}
\caption{Panel A: Estimated ungrouped death rates over single year of age, calendar year, and week. Bottom panels: Observed death rates  by 5-year age groups alongside estimated ungrouped mortality rates. Panel B: Week 1 in 2010 over age. Panel C: Ages 70-74 in Week 1 over time. Panel D: Ages 70-74 in Year 2010 over weeks. Areas depicted in Panel A identify the age-year-week combinations plotted in the three bottom panels.}
\label{fig:p_Ages_Years_Weeks}
\end{figure}

For the estimation, we employed a full interaction model to account for the complex interplay between age, years, and weeks in the mortality data. The model's configuration utilized $(c_1, c_2, c_3)=(21,4,10)$ $B$-splines to capture smooth variations over age, years, and weeks, respectively. Penalization was applied using second-order differences across all three dimensions, with smoothing parameters set to $(\lambda_1,\lambda_2, \lambda_3)=c(30,0.1,100)$. With these settings, the fitted model has effective degrees of freedom of 134. 

The outcomes of the model are presented in Figure \ref{fig:p_Ages_Years_Weeks}. The top panel (A) illustrates the smooth and ungrouped (over age) log-mortality surfaces across age, years, and weeks, providing a comprehensive view of the mortality dynamics. The three bottom panels offer cross-sectional visualizations of the top panel, focusing on specific dimensions. These include observed and estimated log-mortality over ages for a given year and week (B), over years for a specific age group and week (C), and over weeks for a specific age group and year (D). To convey the uncertainty associated with the estimates, 95\% confidence intervals are displayed alongside the results. The bottom panels underline the model's ability to provide a smooth and detailed description of mortality dynamics across all dimensions while simultaneously disaggregating trends to single years of age. 

The theoretical full model matrix, $\bm{B}=\bm{B}_{3} \otimes \bm{B}_{2} \otimes \bm{B}_{1}$, required for estimating a PCLM in three dimensions, would have dimensions of $(m_{1}m_{2}m_{3} \times c_{1}c_{2}c_{3})=(109,200 \times 840)$. Similarly, the composition matrix $\bm{C}$, essential for the original estimation algorithm, would have dimensions of $(n_{1}n_{2}n_{3} \times m_{1}m_{2}m_{3})=(19,760 \times 109,200)$. Both matrices are computationally prohibitive to construct or handle on standard personal computers due to their enormous size. 

With our approach, these matrices do not need to be explicitly constructed. This eliminates the associated storage and computational challenges, allowing the model to be fitted efficiently and without memory constraints. Notably, the combined size of all relevant \texttt{R} objects, including the data and resulting outcomes, amounts to only 319 MB.

Despite the model's complexity, as indicated by the large number of penalized parameters ($\bm{\alpha}$, equal to $c_{1}c_{2}c_{3}=840$), the adoption of the GLAM arithmetic enables efficient estimation within a reasonable computational time: approximately 25 seconds under the model settings and the previously described hardware configuration. Notably, in this three-dimensional case, a substantial portion of this time (45\%) is dedicated to computing the model uncertainty, a step where the full GLAM algorithm cannot be utilized (see Section \ref{sec:uncertainty}). This percentage is notably higher than in the two-dimensional case, reflecting the increased computational demands of the higher-dimensional model.

This computational efficiency not only highlights the robustness of our approach but also underscores its critical role when dealing with grouped data. In cases where observations are provided across multiple dimensions, such as age, year, and week, and are potentially aggregated over one or more dimensions, the original algorithm cannot handle the resulting high-dimensional structure. Our advancements are therefore essential for enabling the estimation of complex PCLMs in such scenarios, offering a practical and scalable solution for analyzing grouped datasets.

\section{Conclusions}\label{sec:Conclusions}

In fields like demography, epidemiology, or economics, data aggregation over different dimensions such as age or time is often unavoidable due to privacy concerns or data collection constraints. Such aggregation can obscure latent patterns that are crucial for understanding underlying processes and informing policy decisions. The penalized version of the Composite Link Model model enables the disaggregation of grouped data to capture fine-grained trends in mortality and other datasets, offering practical utility for applications requiring detailed insights. However, model estimation becomes computationally demanding and sometimes even impossible, since memory and processing demands can become prohibitive when the dimension of the array and/or the desired refinement of the latent distribution to be estimated increases.

To address this issue, we proposed a modified version of the original algorithm introduced by \cite{ThompsonBakerCLM1981}, reformulating the PCLM estimation process. This reformulation is based on defining a working latent response rather than a working regression matrix. This is similar to the E-step in the EM algorithm, but is done purely for computational convenience. Unlike the EM algorithm, which assumes a distribution for the latent data, our approach assumes a Poisson distribution for the observed aggregated counts. This key distinction ensures that the method focuses directly on the observed data without alternating between observed and latent distributions. By redefining the algorithm in terms of the working response, the entire process is reformulated as a Generalized Linear Array Model \citep[GLAM,][]{CurrieDurbanEilersGLAM2006}, significantly reducing computational complexity and enabling the estimation of models that were previously computationally infeasible. 

The simulation studies and real-world applications presented in this paper demonstrate the versatility and computational efficiency of the proposed methodology in the context of grouped count data. The first simulation study reaffirmed the high accuracy of the PCLM in recovering the latent distribution from aggregated data. The second study addressed the central contribution of this work, evaluating the performance of the new GLAM-based estimation algorithm across a variety of data structures, including two-, three-, and four-dimensional settings. The results clearly demonstrate that the proposed approach achieves substantial computational gains, by orders of magnitude, compared to the original PCLM algorithm; in fact, for large datasets, which are common in empirical applications, the original algorithm often fails to produce estimates altogether, making the proposed method not only more efficient but in many cases the only viable solution.

These findings are further supported by real-world applications. In particular, Swedish female mortality grouped by age and year is analyzed, illustrating the method's capability to disaggregate aggregated counts into finer resolutions while preserving accuracy and computational feasibility. A more complex dataset of Spanish male mortality, classified by age group, year, and week, highlights the methodology's extension to three-dimensional settings. These applications underscore the practicality of the approach for handling large-scale, multidimensional grouped observations, demonstrating its potential for generating fine-grained insights from aggregated observations.

Future research could explore several avenues to further improve the applicability and efficiency of the proposed methodology. One promising direction is extending the approach to handle more complex, high-dimensional datasets, especially those involving additional covariates. While the current method is effective for two- and three-dimensional cases, scaling it to higher-dimensional settings, such as spatial-temporal data or multi-level hierarchical structures, presents challenges, particularly since the use of GLAM may not be feasible when the model matrix cannot be expressed as Kronecker products.
Another potential avenue is the incorporation of alternative probabilistic models for the latent variables, moving beyond the Poisson distribution, which could provide greater flexibility for handling diverse types of aggregated count data.

\bigskip\bigskip\bigskip
\subsubsection*{Acknowledgments}
The authors would like to express their gratitude to Paul Eilers for his insight on some aspects of this work. This paper is part of the grant  PID2022-137243OB-I00 funded by MCIN/AEI/ 10.13039/501100011033 and,
by ``ERDF A way of making Europe''. This support is gratefully acknowledged.

\newpage
\appendix

\section{Computational Implementation}\label{sec:AppAComputation}
As mentioned in the main body of the document, \cite{CurrieDurbanEilersGLAM2006} developed an arithmetic for arrays that enables high-speed computation in the scoring algorithm of generalized linear models. This arithmetic is based on a sequence of nested matrix operations, such that by reorganizing the computations, it becomes possible to perform operations using arrays of the same size as the data. This approach avoids the need to vectorize the data on the array or flatten the tensor products to a regression basis, resulting in significant computational gains as the array's dimensionality increases. Here, We outline the fundamental operations required to solve \eqref{eq:newIRLS} and provide their implementation in \texttt{R} code. The fully reproducible program, which estimates the models for both datasets and reproduces the plots presented in this paper, is available at \url{osf.io/uwejt/?view_only=2ca1fdb7568342bbb9a3c51fd33c718c}.

\subsubsection*{Row tensor}

The row tensor of  matrices $\bm{X}_1 $ and $\bm{X}_2$ of dimensions $n \times c_1$ and  $n\times c_2$, respectively is defined as:
\begin{equation*}
\mathcal{G}(\bm{X}_1,\bm{X}_2) = (\bm{X}_1\otimes \bm{1}_{c_2}') \odot (\bm{1}_{c_1}' \otimes \bm{X}_2)
\end{equation*}
The operation described above  is such that row $i$ of $\mathcal{G}(\bm{X}_1,\bm{X}_2)$ is the Kronecker product of row $i$  of $\bm{X}_1$ by row $i$  of $\bm{X}_2$.

In \texttt{R} this function translates to:
\singlespacing
\begin{lstlisting}[language=R]
Rten <- function(X1,X2){
	one.1 <- t(rep(1,ncol(X1)))
	one.2 <- t(rep(1,ncol(X2)))
	kronecker(X1, one.1) * kronecker(one.2, X2)
}\end{lstlisting}
\onehalfspacing

\subsubsection*{$\mathcal{H}$-transform}

The $\mathcal{H}$-transform generalizes to $d$-dimensional arrays the premultiplication of vectors and matrices by a matrix. The $\mathcal{H}$-transform of the $d$-dimensional array $\bm{A}$ of size $c_1\times c_2 \times \cdots \times c_d$ by the matrix $\bm{X}$ of size $r\times c_1$ is denoted $\mathcal{H}(\bm{X},\bm{A})$. If $\bm{A}$ is a vector $\bm{a}$,  $\mathcal{H}(\bm{X},\bm{a})=\bm{X}\bm{a}$, while if $\bm{A}$ is a matrix, $\mathcal{H}(\bm{X},\bm{A})=\bm{X}\bm{A}$. In the case of a  $d$-dimensional, array the premultiplication is carried out as follows: let $\bm{A}^{*}$ of size $c_1\times c_2 c_3 \ldots c_d$ the matrix obtained by flattering dimensions 2 to $d$ of $\bm{A}$; form the matrix product $\bm{X}\bm{A}^{*}$ of size $r\times c_2c_3\ldots c_d$; then $\mathcal{H}(\bm{X},\bm{A})$ is the $d$-dimensional array of size $r\times c_2\times \cdots \times c_d$ obtained from $\bm{X}\bm{A}^{*}$ by reinstating dimensions 2 to $d$ of $\bm{A}$. 

In \texttt{R}, the $\mathcal{H}$-transform can be implemented as follows:
\singlespacing
\begin{lstlisting}[language=R]
H <- function(X, A){
	d <- dim(A)
	M <- matrix(A, nrow = d[1])
	XM <- X %*% M
	array(XM, c(nrow(XM), d[-1]))
}
\end{lstlisting}
\onehalfspacing

\subsubsection*{Rotation of a $d$-dimensional array}

The rotation of the $d$-dimensional array $\bm{A}$ of size $c_1\times c_2 \cdots c_d$ is the $d$-dimensional array $R(\bm{A})$ of size $c_2\times c_3 \cdots c_d \times c_1$ obtained by permuting the indices of $\bm{A}$.

\singlespacing
\begin{lstlisting}[language=R]
Rotate = function(A){
    d = 1:length(dim(A))
    d1 = c(d[-1], d[1])
    aperm(A, d1)
}
\end{lstlisting}
\onehalfspacing

\subsubsection*{Rotated $\mathcal{H}$-transform}

The rotated $\mathcal{H}$-transform of the array $\bm{A}$ by the matrix $\bm{X}$ is given by
\begin{equation*}
	\label{eq:CH2a_RotatedTransform}
	\rho(\bm{X},\bm{A}) = \mathcal{R}(\mathcal{H}(\bm{X},\bm{A}))
\end{equation*}
with the associated \texttt{R} function:
\singlespacing
\begin{lstlisting}[language=R]
RH <- function(X, A){
    Rotate(H(X, A))
}
\end{lstlisting}
\onehalfspacing

\clearpage
\bibliographystyle{chicago}
\bibliography{Bibliography}

\end{document}